\pdfoutput=1
\documentclass[a4paper,11pt]{article}
\usepackage{jheppub}

\usepackage[utf8]{inputenc}
\usepackage[T1]{fontenc}

\usepackage{lmodern}
\usepackage{microtype}
\usepackage{amsmath,amsfonts,amssymb,amsthm}
\usepackage{mathrsfs}

\usepackage{cleveref}
\usepackage{graphicx}
\usepackage{tensor}
\usepackage{braket}
\usepackage{mleftright} \mleftright
\usepackage{comment}
\usepackage{array}
\usepackage{booktabs}
\usepackage{todonotes}
\usepackage{mathtools} 
\usepackage{color}
\usepackage{bbold}
\allowdisplaybreaks

\crefname{equation}{}{}

\DeclareMathOperator{\extdm}{d}

\newcommand{\extd}{\extdm \!}
\newcommand*{\dd}{\extd}

\newcommand*{\R}{{\mathbb{R}}}

\newcommand*{\Z}{{\mathbb{Z}}}

\providecommand{\Lt}{{\tt L}}
\renewcommand{\Lt}{{\tt L}}

\newcommand{\eq}[2]{\begin{equation} #1 \label{#2} \end{equation}}

\begin{document}

\preprint{IPM/P-2019/040}
\rightline{TUW--19--02}

\title{Interpolating Between Asymptotic and Near Horizon Symmetries}

\author[\heartsuit]{Daniel Grumiller,}
\author[\diamondsuit\spadesuit]{M.M. Sheikh-Jabbari,}
\author[\clubsuit]{Cedric Troessaert,}
\author[\heartsuit]{and Raphaela Wutte}

\affiliation[\heartsuit]{Institute for Theoretical Physics, TU Wien, Wiedner Hauptstrasse 8--10/136, A-1040 Vienna, Austria}
\affiliation[\diamondsuit]{School of physics, Institute for research in fundamental sciences (IPM), P.O.Box 19395-5531, Tehran, Iran}
\affiliation[\spadesuit]{The Abdus Salam ICTP, Strada Costiera 11, 34151, Trieste Italy}
\affiliation[\clubsuit]{Haute-Ecole Robert Schuman, Rue Fontaine aux M\^ures, 13b, B-6800 Libramont, Belgium}

 \emailAdd{grumil@hep.itp.tuwien.ac.at, jabbari@theory.ipm.ac.ir, cedric.troessaert@hers.be, rwutte@hep.itp.tuwien.ac.at}


\abstract{
We develop basic tools and matching conditions to interpolate between asymptotic and near horizon symmetries. We focus on black holes in three dimensions. In particular, we match Brown--Henneaux boundary conditions at infinity, which yields two Virasoro algebras, to Heisenberg boundary conditions at the horizon yielding two $\hat u(1)$ current algebras. Our construction allows to equip  BTZ black holes with soft hair excitations at the horizon invisible to the asymptotic observer. 
}

\maketitle
\flushbottom

\section{Introduction}\label{se:0}

It has been advocated for a long time that generic (non-extremal) black hole horizons are special loci in spacetime and that it could make sense to impose boundary conditions at these loci \cite{Strominger:1997eq, Carlip:1998wz, Carlip:1999cy}. In the past couple of years this idea has been implemented in various ways, leading to different near horizon symmetries \cite{Donnay:2015abr, Afshar:2015wjm, Afshar:2016wfy, Carlip:2017xne, Haco:2018ske, Chandrasekhar:1985kt, Grumiller:2019fmp, Aggarwal:2019iay, Carlip:2019dbu, toappear:2019}. In nearly all the cases the near horizon symmetries differ from the usual asymptotic symmetries. It is thus fair to ask how the former match with the latter and if we can interpolate between them. This question provides the main motivation for the present work. To address possible interpolations between asymptotic and near horizon symmetries we focus on three-dimensional black holes, since this is the cleanest situation where this issue arises.

From an asymptotic viewpoint the most natural set of asymptotic symmetries, the two-dimensional conformal algebra, was derived by Brown and Henneaux \cite{Brown:1986nw}. From a near horizon viewpoint a natural set of near horizon symmetries consists of two copies of $\hat u(1)$ current algebras (or, equivalently, of Heisenberg algebras) \cite{Afshar:2016wfy}. If one naively extrapolates either of these boundary conditions, respectively, to the other boundary one does not recover the appropriate set of symmetries. For instance, extrapolating the Heisenberg boundary conditions all the way from the horizon to infinity establishes asymptotic boundary conditions that differ from the ones of Brown--Henneaux \cite{Afshar:2016uax}. Similarly, naively moving the cutoff surface where Brown--Henneaux boundary conditions are imposed all the way from infinity to the horizon maintains the conformal symmetries and does not lead to Heisenberg boundary conditions at the horizon.\footnote{In the two cases discussed here (Heisenberg and Brown--Henneaux), the respective symmetries are symplectic and not just asymptotic \cite{Compere:2015knw}. This means that one can construct a family of locally AdS$_3$ solutions which form a phase space, as has been done respectively in \cite{Afshar:2015wjm, Afshar:2016wfy} and \cite{Banados:1998gg}. Given this phase space one may then define the symmetries away from where the initial boundary conditions have been imposed.}

In this work we clarify how a smooth matching works that connects arbitrary (consistent) sets of boundary conditions at two different boundaries. Phrased differently, our paper provides a manual how to grow soft hair (in the sense of Hawking, Perry and Strominger \cite{Hawking:2016msc}) on black holes that is invisible to asymptotic observers. While for concrete applications we imagine one of these boundaries to be the asymptotic one and the other the (stretched) horizon, our results are general and apply to any sets of two disconnected boundary components. 

A related gravitational motivation is something one could call ``causal patch holography'', where the boundary considered is a causal patch in some Penrose diagram (e.g.~the diamond-shaped region between two horizons or the region between the event horizon and the asymptotic boundary). Introducing a finite cutoff at both boundaries renders the cutoff surfaces timelike in most applications, so that we are in a situation where we have two disconnected boundary components, e.g.~one at a radial distance close to the black hole horizon and the other one close to the asymptotic boundary. Since we are going to employ the Chern--Simons formulation of three-dimensional gravity \cite{Achucarro:1987vz, Witten:1988hc} our results also have potential applications beyond gravity. 

This paper is organized as follows.  In section \ref{se:1} we review salient features of Chern--Simons theories.  In section \ref{se:2} we discuss Chern--Simons theories with two boundaries and address general aspects of the matching conditions. In section \ref{se:3} we analyze the prime example, connecting asymptotic Brown--Henneaux with near horizon Heisenberg boundary conditions. In section \ref{se:4} we treat another example, connecting asymptotic Comp{\`e}re--Song--Strominger with near horizon Heisenberg boundary conditions. In section \ref{se:5} we conclude with a survey of further applications and generalizations. 
\section{Chern--Simons}\label{se:1}

We recall now salient features of Chern--Simons theories, without yet emphasizing their specific gravitational r\^ole.
Three dimensional Chern--Simons  gauge theories with the Lie-algebra valued gauge connection $A$ are governed by the action
\begin{equation}
\label{Chern--Simons action}
I_{\textrm{\tiny CS}} [A] = \frac{k}{4 \pi}  \int \braket{ A \wedge \dd A + \frac{2}{3} A \wedge A \wedge A }\,.
\end{equation}
The only coupling constant is the dimensionless Chern--Simons level $k$, and the brackets $\braket{\,, \,}$ denote an invariant, non-degenerate bilinear form on the gauge algebra. In a derivative expansion the three-dimensional, parity-odd, Chern--Simons term \eqref{Chern--Simons action} is the lowest derivative term leading to gauge-covariant equations of motion
\begin{equation}
\label{EOM CS}
F = \dd A + A \wedge A = 0\,.
\end{equation}
Chern--Simons theories arise universally in three spacetime dimensions. Prominent examples are quantum Hall systems \cite{Zhang:1988wy, Susskind:2001fb, Polychronakos:2001mi} and three-dimensional gravity \cite{Deser:1982vy, Deser:1982wh, Achucarro:1986vz, Witten:1988hc, Carlip:1995zj, Carlip:1998uc}. Locally the equations of motion \eqref{EOM CS} imply gauge flatness of the connection,
\begin{equation}
\label{trivial gauge}
A = g^{-1} \dd g
\end{equation}
where $g$ is some group element of the underlying gauge group. This means there are no local physical degrees of freedom in the Chern--Simons theory; it is a topological quantum field theory of Schwartz type \cite{Birmingham:1991ty}. 

Despite of local triviality, systems described by Chern--Simons theories on manifolds with boundaries can have physical degrees of freedom, often referred to as ``edge-states'' in quantum Hall contexts (see e.g.~\cite{Balachandran:1991dw}) and ``boundary gravitons'' in gravitational contexts (see \cite{Brown:1986nw} for the seminal construction). Boundary conditions on the gauge connection $A$ are a crucial physical input in the theory. Different choices are possible and can lead to different physical phase spaces and symmetry algebras acting on them, see \cite{Grumiller:2016pqb} for a summary of possibilities in three-dimensional Einstein gravity with negative cosmological constant. 

In all the constructions and applications mentioned above there is a single (actual or asymptotic) simply connected boundary (``on which the edge excitations live''). In the next sections we consider Chern--Simons theories on manifolds with two boundaries.

\section{Chern--Simons with two boundaries}\label{se:2}

The goal of this section is to determine whether it is possible to consistently impose two different sets of falloff conditions on the Chern--Simons connection at the two boundaries.

For concreteness we focus on the example of three-dimensional Euclidean Einstein gravity with negative cosmological constant in the Chern--Simons formulation. We consider a manifold with two boundaries with the topology of a torus, $S^1 \times S^1$. We use the coordinate system $\tau, r, \varphi$, where $r$ is the radial coordinate, $\varphi\sim\varphi+2\pi$ is the angular coordinate and $\tau \sim \tau + \beta$ corresponds to the Euclidean time. One boundary is located at $r = 0$, while the other one is located at $r \to \infty$. The cycles of the torus are identified through
    $(\tau, \varphi) \sim (\tau, \varphi + 2 \pi) \sim (\tau + \beta, \varphi )$,  
where $\beta$ is the inverse temperature.\footnote{We could have considered the angular velocity $\Omega$ in the identification $(\tau,\,\varphi)\sim (\tau+\beta,\,\varphi +\beta\Omega)$. However, the information about rotation can also be packaged into the values of the chemical potentials. So, we always have the same torus regardless of the value of angular velocity. See e.g.~\cite{Compere:2013nba, Henneaux:2013dra}.} The $\tau$-cycle is contractible, while the $\varphi$-cycle is non-contractible. 

In section \ref{se:2.1} we quickly summarize the main aspects of Euclidean AdS$_3$ Einstein gravity as Chern--Simons theory. In section \ref{se:2.2} we show how to interpolate the gauge connection between two boundaries. In section \ref{se:2.3} we derive matching conditions that relate some of the charges at the two boundaries.

\subsection{\texorpdfstring{AdS$_3$}{AdS3} Einstein gravity as Chern--Simons theory}\label{se:2.1}

In the Chern--Simons formulation of AdS$_3$ Einstein gravity \cite{Achucarro:1987vz, Witten:1988hc} the Chern--Simons level $k$ in the action \eqref{Chern--Simons action} is (one quarter of) the ratio between AdS-radius $\ell$ and Newton's constant $G_N$.
\begin{equation}
    k = \frac{\ell}{4 G_N}
\end{equation}
In the remainder of this work we set $\ell=1$. The gauge algebra for AdS$_3$ Einstein gravity is $sl(2, \R) \oplus  sl(2, \R)$. This split can also be performed on the level of the action,
\begin{equation}
\label{ics}
I_{\textrm{\tiny CS}}[A^+, A^-] = I_{\textrm{\tiny CS}}[A^+] -I_{\textrm{\tiny CS}}[A^-]
\end{equation}
where $A^+$ and $A^-$ are elements of $sl(2, \R)$ each. Since $A^+$ and $A^-$ decouple, we restrict our discussion 
mostly to one $sl(2, \R)$ sector in the following and to reduce clutter drop the superscript plus. Hereafter,  $A$ without superscript refers to $A^+$. We use the standard $sl(2,\R)$ generators $\Lt_1, \Lt_0$ and $\Lt_{-1}$ with commutation relations $[\Lt_n,\,\Lt_m]=(n-m)\,\Lt_{n+m}$. The non-degenerate, invariant bilinear form on the algebra is given by $\braket{L_0, L_0}=1/2$, $\braket{L_1, L_{-1}}= \braket{L_{-1}, L_{1}} = -1$ and  $\braket{L_n, L_m}= 0$ otherwise.

\subsection{Gauge connection in presence of two boundaries}\label{se:2.2}

To find a flat connection interpolating between a certain set of boundary conditions at $r= 0$ and a (possibly different) set of boundary conditions at $r \to \infty$ we make the following ansatz. We take a solution to the field equations \eqref{EOM CS} subject to \emph{one} set of boundary conditions at $r \to \infty$, $A_\infty$, and a solution subject to \emph{a different} set of boundary conditions at the boundary $r = 0$, $A_0$. Next we rewrite both of them in the form \eqref{trivial gauge}. In general, this is impossible with single-valued group elements $g$. However, we allow for multivaluedness of the group element $g$ and thus continue to employ \eqref{trivial gauge} even globally.

Using the Gauss decomposition, we write the group element $g_\infty$ for, say, the solutions $A_\infty$ as\footnote{In addition to the Gauss decomposition we have pulled out an extra factor $e^{r \Lt_0}$, which is common practice in order to make the functions $\eta_\infty$, $\psi_\infty$ and $\lambda_\infty$ $r$-independent. While in our matching the corresponding functions will be $r$-dependent, we still pull out such a factor in order to asymptote to well-known expressions.}
\begin{equation}
\label{groupelementinfty}
g_\infty = e^{\eta_\infty \Lt_1}\, e^{\psi_\infty \Lt_0}\, e^{\lambda_\infty \Lt_{-1}}\, e^{r \Lt_0}\,.
\end{equation}
The functions $\eta_\infty = \eta_\infty(\tau, \varphi)$, $\psi_\infty = \psi_\infty(\tau, \varphi)$ and $\lambda_\infty = \lambda_\infty(\tau, \varphi)$ are allowed to be multivalued. The connection \eqref{trivial gauge} associated to the group element \eqref{groupelementinfty} can be rewritten as
\eq{
	A_\infty = b^{-1}\,(\dd + a_\infty)\,b\,, \qquad\qquad b= e^{r \Lt_0}
}{eq:whatever}
with
\eq{
	a_\infty = e^{\psi_\infty} \dd\eta_\infty\, \Lt_1 + (\dd\psi_\infty + 2 \lambda_\infty e^{\psi_\infty} \dd\eta_\infty)\,
	\Lt_0 + 
	(\dd\lambda_\infty + \lambda_\infty \dd\psi_\infty + \lambda^2_\infty e^{\psi_\infty} \dd\eta_\infty)\, \Lt_{-1}\,.
}{connectioneq}
The connection $A_0$ is decomposed in the same way, replacing everywhere subscripts $\infty$ by $0$.

To obtain the full connection $A$ we use a suitable profile function to smoothly interpolate between the connection $A_0$, parametrized by $\eta_0, \psi_0$ and $\lambda_0$ and the connection $A_\infty$, parametrized by $\eta_\infty, \psi_\infty$ and $\lambda_\infty$. A group element $g$ interpolating between the two boundaries can be built by first Gauss decomposing as above
\begin{equation}
\label{groupelement}
g = e^{\eta \Lt_1}\, e^{\psi \Lt_0}\, e^{\lambda \Lt_{-1}}\, e^{r \Lt_0}
\end{equation}
with
\begin{equation}
\label{interpolel}
\eta = \frac{\eta_0 + \eta_\infty f(r)}{1 + f(r)}\,, \qquad\qquad \lambda = \frac{\lambda_0 + \lambda_\infty f(r)}{1 + f(r)}\,, \qquad\qquad \psi = \frac{\psi_0 + \psi_\infty f(r)}{1 + f(r)}\,,
\end{equation}
such that the connection $A$ that interpolates between the two boundaries is given by
\eq{
	A = b^{-1}\,\big(\dd + a\big)\, b\,, \qquad\qquad b= e^{r \Lt_0}
}{eq:2b1}
with
\eq{
	a = e^{\psi} \dd\eta\, \Lt_1 + \big(\dd\psi + 2 \lambda e^{\psi} \dd\eta\big)\,
	\Lt_0 + 
	\big(\dd\lambda + \lambda \dd\psi + \lambda^2 e^{\psi} \dd\eta\big)\, \Lt_{-1}\,.
}{eq:2b2}

The information of the specific boundary conditions one tries to match will be encoded in the interpolating function $f(r)$. Even for a given set of boundary conditions, there is a large (pure gauge) ambiguity in how to choose this function. The minimal requirements are $\lim_{r\to\infty} f(r)\to\infty$ and $\lim_{r\to 0}f(r)\to 0$, ensuring that we match with $A_\infty$ and $A_0$, respectively. Moreover, we demand that $f(r)$ is monotonically increasing as we move from $r=0$ to $r=\infty$.

In order to satisfy the desired sets of boundary conditions at either of the boundaries the interpolating function $f(r)$ additionally should have appropriate fall-off behavior at both boundaries to make sure that the modifications to the connection produced by our gluing are subleading, i.e., pure gauge from either boundary theory point of view. As we shall see in the prime example in section \ref{se:3} it is simple to find such functions $f(r)$.

\subsection{Matching conditions}\label{se:2.3}

The group element $g$ in \eqref{groupelement}-\eqref{interpolel} corresponds to a connection \eqref{trivial gauge} satisfying the equations of motion \eqref{EOM CS} and both sets of asymptotic conditions. However, we are not done yet. We need to make sure that by interpolating between $A_\infty$ and $A_0$ we do not introduce global inconsistencies. One simple example of an inconsistency is a configuration $A_\infty$ that requires Euclidean time $\tau$ to have a certain periodicity, $\tau\sim\tau+\beta_\infty$, while $A_0$ requires also some periodicity, $\tau\sim\tau+\beta_0$. Unless both periodicities coincide, $\beta_\infty=\beta_0$, no smooth matching between $A_\infty$ and $A_0$ is possible. In physical terms this means we have to make sure that corresponding temperatures (and possibly other quantities) match when interpolating. 

More generally, we have to compare holonomies calculated at infinity with holonomies calculated at zero and demand that the results agree with each other. If they did not agree this would imply that we introduced sources or singularities somewhere in the middle of the manifold, which needs to be avoided for a smooth matching. 
In our specific setup, that amounts to imposing that the holonomy around the $\tau$-cycle is trivial and that the holonomies around the $\varphi$-cycle coincide for the connections we want to match. In what follows we carry this out by introducing a group element $\tilde{h}$ based on the holonomy.


The group element $g$ and the Wilson line\footnote{As usual, $\mathcal{P}$ denotes path ordering.}
\begin{equation}
h(r, \tau, \tau_0,\varphi) = \mathcal{P} \exp  \int\limits_{\tau_0}^{\tau} A_{\tau}(r, \tau', \varphi) \extd \tau'
\end{equation}
fulfill the same differential equation,
\eq{
A_\tau(r,\tau, \varphi) = g^{-1}(r, \tau, \varphi)  \partial_\tau g(r,\tau, \varphi)\,,\quad A_\tau(r, \tau, \varphi)  = h^{-1}(r, \tau, \tau_0, \varphi)  \partial_\tau h(r, \tau, \tau_0, \varphi) .
}{m1}
Using that  $h(r, \tau_0, \tau_0, \varphi) = \mathbb{1}$ and that the group element $g$ solves the differential equation \eqref{m1} on a constant $\varphi$ slice yields
\begin{equation}
    h(r,\tau, \tau_0,\varphi)  = g^{-1}(r, \tau_0,\varphi) g(r, \tau,\varphi)\,.
\end{equation}
{From the holonomy $h(r,\tau + \beta, \tau, \varphi)$, we can build a new group element $\tilde h(r,\tau + \beta, \tau, \varphi)$,
\begin{equation}
\label{ceq}
    \tilde h(r,\tau+\beta, \tau, \varphi) = g(r,\tau,\varphi)h(r,\tau + \beta, \tau, \varphi)g^{-1}(r, \tau, \varphi)
    = g(r, \tau + \beta, \varphi) g^{-1}(r, \tau, \varphi),
\end{equation}
which corresponds to a Wilson line of an auxiliary connection $gA_{\tau} g^{-1}$, as
\begin{align}
    \partial_\tau \tilde h(r, \tau, \tau_0, \varphi) &= (g A_\tau g^{-1})(r, \tau, \varphi)\, \tilde h(r, \tau, \tau_0, \varphi)\,, \\
    \tilde h^{-1}(r, \tau, \tau_0, \varphi) &= \mathcal P \exp \int \limits_{\tau_0}^{\tau} (- gA_{\tau} g^{-1})(r, \tau', \varphi) \extd \tau'\,.
\end{align}}
This group element can be used to control the periodicity. In particular, one can show that the connection is periodic of period $\beta$ along the $\tau$-cycle iff $\tilde h(r,\tau + \beta, \tau, \varphi)$ is constant, i.e. iff the following three expressions are constant in $r$, $\tau$ and $\varphi$
\begin{gather}
    e^{\frac{\psi_h}{2}} = e^{\frac{\Delta\psi}{2}} \big(1 + \Delta\lambda \, \eta e^{\psi}\big), \qquad
    \lambda_h e^{\frac{\psi_h}{2}} = \Delta\lambda e^{\psi + \frac{\Delta\psi}{2}}, \nonumber\\
    \eta_h e^{\frac{\psi_h}{2}} = e^{\frac{\Delta\psi}{2}} \big[\Delta\eta+ \eta(1-e^{-\Delta\psi})+ \eta \Delta\lambda (\eta+\Delta\eta)e^\psi\big],
    \label{eq:ced1}
\end{gather}
where $\Delta X\equiv X(\tau+\beta)-X(\tau)$ and $X_h$ correspond to the Gauss decomposition of $\tilde h$.

{
The triviality of the $\tau$-cycle further requires $h(r, \tau_0 + \beta,\tau_0, \varphi) = \tilde h(r, \tau_0 + \beta, \tau_0, \varphi) = \pm \mathbb{1} $ leading to $\Delta\lambda=\Delta \eta=0, e^{\Delta\psi}=1$, or equivalently,
\begin{align}
\label{lz18}
\eta(r,\tau + \beta,\varphi) &= \eta(r,\tau,\varphi), & \lambda(r,\tau
+ \beta,\varphi)& = \lambda(r,\tau,\varphi),\nonumber \\
\psi(r,\tau +
\beta,\varphi) &= \psi(r,\tau,\varphi) + 2 \pi i\,n,\ \quad n\in \mathbb{Z}& & &
\end{align}
In particular, these conditions must hold for
$\eta_\infty$, $\psi_\infty$ and $\lambda_\infty$ and $\eta_0$, $\psi_0$ and
$\lambda_0$ for each constant $\varphi$ slice. With this, we have shown that the matching of the periodicity in $\tau$ of
the two asymptotic solutions is necessary for
the regularity in $\tau$ of the new connection. One can easily check that, due to the form \eqref{groupelement} of the
interpolating group element $g$, this is also sufficient.}

{We have dealt with the $\tau$-cycle and focus now on the $\varphi$-cycle. The
 group element controlling the periodicity in $\varphi$ takes the form
\begin{equation}
    \tilde{h}_\varphi (r,\tau,\varphi + 2 \pi, \varphi) = g(r, \tau, \varphi + 2 \pi) g^{-1}(r, \tau, \varphi).
\end{equation}
The periodicity in $\varphi$ is equivalent to the expressions in \eqref{eq:ced1} being constant but, this time, $\Delta X= X(\varphi + 2\pi)- X(\varphi)$. In particular, when evaluating this expression for both asymptotic group
elements $g_0$ and $g_\infty$, we must obtain the same value.}

As a consequence of matching the holonomies $\tilde h$, the functions $\eta_\infty$, $\psi_\infty$, $\lambda_\infty$ are not completely independent from the functions $\eta_0$, $\psi_0$, $\lambda_0$. We shall see how the matching conditions work in detail when discussing examples. We start with the first one in the next section.

\section{Brown--Henneaux/Heisenberg matching}\label{se:3}

We discuss now our prime example, the matching between Brown--Henneaux boundary conditions at infinity and Heisenberg boundary conditions at the horizon. 

In section \ref{se:3.1} we present the connection at both boundaries. In section \ref{se:3.2} we study consequences of the matching conditions. In section \ref{se:3.3} we collect the results and present the full gauge connection that smoothly interpolates {between Brown Henneaux boundary conditions at infinity and Heisenberg boundary conditions at $r= 0$}
as well as the corresponding metric. In section \ref{se:3.4} we show how to interpolate non-smoothly using the Heaviside function while maintaining smoothness of curvature.

\subsection{Connection in presence of two boundaries}\label{se:3.1}
\label{conn2boundBH} 

Consider Brown--Henneaux boundary conditions at $r\to \infty$ and Heisenberg boundary conditions at $r = 0$. Our first task is to bring the connection into the form used in section \ref{se:2.2}.

For a theory  with just one boundary at $r \to \infty$, the connection subject to the Brown--Henneaux boundary conditions with constant chemical potentials  $\mu$ in the plus-sector, see \eqref{ics}, in Euclidean time is given by \cite{Brown:1986nw, Banados:1998gg, Carlip:1998uc}
\eq{
A_{\infty} = b^{-1}\, \big( \dd + a_\infty \big)\, b\,,\qquad\qquad b = e^{r \Lt_0}\,, \qquad\qquad
a_\infty =    \big(\Lt_1 - \frac{1}{2} \mathcal{L} L_{-1}\big) \dd z\,,
}{Ainfty}
where $z = i \mu \tau + \varphi$ and $\bar{z} = -i \mu \tau + \varphi$.
On-shell $ \mathcal{L} = \mathcal{L}(z, \bar{z})$ obeys
\begin{equation}
\label{onshell}
    \partial_{\bar{z}} \mathcal{L} = 0\,.
\end{equation}
 As any holomorphic function on a compact {connected} complex manifold (such as a torus) must be constant, $\mathcal{L}$ is constant.
{In the following, we assume without loss of generality that $\mathcal{L}$ is positive.}

Alternatively, for a theory with just one boundary at $r = 0$, imposing Heisenberg boundary conditions with constant chemical potentials \cite{Afshar:2016wfy}, the connection can be written as
\eq{
A_0 = b^{-1}\, \big( \dd + a_0 \big)\, b\,, \qquad b =  e^{r \Lt_0}\,, \qquad a_0 = \frac{1}{2} \, \big(\Lt_{-1} - \Lt_{1}\big)\,\big(\mathcal{J}(\varphi)\,\dd \varphi +  i \zeta\,\extd\tau \big)\,,
}{A0new}
where $\zeta = \mathrm{const}$.\footnote{The form of the connection originally provided in \cite{Afshar:2016wfy, Afshar:2016kjj} is recovered by a change of basis: $\tilde{\Lt}_0 = \frac{1}{2} (\Lt_{-1} - \Lt_1)$, $\tilde{\Lt}_1 = \frac{1}{2} (\Lt_1 + 2 \Lt_0 + \Lt_{-1})$, $\tilde{\Lt}_{-1} = \frac{1}{2} (\Lt_1 - 2 \Lt_0 + \Lt_{-1})$.}
We construct now a solution of the field equations subject to the Brown--Henneaux boundary conditions at $r \to \infty$ and subject to the Heisenberg boundary conditions at $r = 0$ along the lines of section \ref{se:2}, without yet specifying the interpolating function $f(r)$. The variables $\eta_{0/\infty}$, $\psi_{0/\infty}$ and  $\lambda_{0/\infty}$ representing the group elements $g_{0/\infty}$ can be chosen as
\begin{subequations}
\label{identlambdainfty}
\begin{align}
 \eta_\infty &= \mp \frac{e^{- \psi_\infty}}{\sqrt{2 \mathcal{L}}}\,, & \psi_\infty &= \pm \sqrt{2 \mathcal{L}} z\,, & \lambda_\infty  &=  \mp \frac{\sqrt{\mathcal{L}}}{\sqrt{2}}\,,  \\
 \eta_0 &= \pm \frac{1}{2} e^{- \psi_0}\,, & \psi_0 &= \pm \int \mathcal{J}(\varphi^\prime) \dd \varphi^\prime \pm i \zeta \tau\,, & \lambda_0 &= \pm 1\, . 
\end{align}
\end{subequations}
These functions obey the following quasi-periodicity conditions
\begin{subequations}
\label{eq:bhnh02}
\begin{align}
\psi_0(\tau,\varphi + 2 \pi) &= \psi_0(\tau,\varphi) + 2 \pi K_0\,, & \psi_\infty(\tau, \varphi + 2 \pi) &= \psi_\infty(\tau, \varphi) + 2 \pi K_\infty\,, \\
\eta_0(\tau, \varphi+2 \pi) &= e^{- 2 \pi K_0} \eta_0(\tau, \varphi)\,, & \eta_\infty(\tau, \varphi+2 \pi) &= e^{- 2 \pi K_\infty} \eta_\infty(\tau, \varphi)\,, \\
\lambda_0(\tau, \varphi + 2 \pi) &= \lambda_0(\tau, \varphi)\,, & \lambda_\infty(\tau, \varphi + 2 \pi) &= \lambda_\infty(\tau, \varphi)\,.
\end{align}
\end{subequations}

Using the Fourier decomposition
\begin{equation}
\label{decomposition}
    J_n = \frac{k}{4 \pi} \int \dd \varphi\, \mathcal{J}(\varphi)\, e^{i n \varphi}\,, \qquad\qquad
    L_0 = \frac{k}{4 \pi} \int \dd \varphi\, \mathcal{L}=\frac{k}{2}\mathcal{L}
\end{equation}
we find $K_\infty = \pm \sqrt{2 \mathcal{L}}= \pm \sqrt{\frac{4}{k} L_0}$ and $K_0 = \pm \frac{2}{k} J_0$. The quasi-periodicity relations along the $\tau$-cycle are analogous to \eqref{eq:bhnh02},
\begin{subequations}\label{eq:bhnh3}
\begin{align}
\psi_0(\tau + \beta,\varphi) &= \psi_0(\tau,\varphi) + i \beta V_0\,, & \psi_\infty(\tau + \beta, \varphi) &= \psi_\infty(\tau, \varphi) + i \beta V_\infty\,, \\
\eta_0(\tau + \beta, \varphi) &= e^{- i \beta V_0} \eta_0(\tau, \varphi)\,, & \eta_\infty(\tau + \beta, \varphi) &= e^{- i \beta V_\infty} \eta_\infty(\tau, \varphi)\,, \\
\lambda_0(\tau + \beta, \varphi) &= \lambda_0(\tau, \varphi)\,, & \lambda_\infty(\tau + \beta, \varphi) &= \lambda_\infty(\tau, \varphi)\,,
\end{align}
\end{subequations}
where $V_\infty = \pm \mu \sqrt{2 \mathcal{L}}= \pm \mu \sqrt{\frac{4}{k} L_0}$ and $V_0 = \pm \zeta$.
The corresponding behaviour for the group element is
\begin{align}
\label{groupperiodicity}
	g_{0/\infty}(\tau, \varphi + 2 \pi) &= e^{2\pi K_{0/\infty} \Lt_0}
	g_{0/\infty}(\tau, \varphi)\,, \\
	g_{0/\infty}(\tau + \beta, \varphi) &= e^{i \beta V_{0/\infty} \Lt_0}
	g_{0/\infty}(\tau, \varphi)\,.
\end{align}

\subsection{Matching conditions}\label{se:3.2}

We study now the matching conditions discussed in section \ref{se:2.3}. We
start with the non-contractible $\varphi$-cycle and consider the corresponding holonomy $\tilde h_\varphi$,
\begin{gather}
    \tilde h_\varphi (r, \tau, \varphi + 2\pi, \varphi) = g(r, \tau, \varphi + 2 \pi) g^{-1}(r, \tau, \varphi)\,. 
\end{gather}
Demanding for consistency that the holonomy $\tilde h_\varphi$ is independent of the coordinates establishes the matching condition
\begin{equation}
K_0 = K_\infty\,.
\end{equation}
While this matching condition requires strict equality between $K_0$ and
$K_\infty$, we have a $\mathbb{Z}_2$ ambiguity in our connections
\eqref{identlambdainfty} which allows us to match solutions with a priori
opposite values of $K_0$ and $K_\infty$. Depending on whether
\begin{equation}
\label{BHphimatch}
    J_0 = + \sqrt{k L_0}\qquad \mathrm{or} \qquad J_0 = - \sqrt{k L_0}
\end{equation}
we choose  
\begin{equation}
    K_0 = \frac{2}{k} J_0\,, \qquad K_\infty = \sqrt{ \frac{4 L_0}{k}} \qquad \mathrm{or} \qquad K_0 = \frac{2}{k} J_0\,, \qquad K_\infty = - \sqrt{ \frac{4 L_0}{k}}\,,
\end{equation}
such that $K_0 = K_\infty$ always. For sake of specificity we consider the case 
\begin{equation}
\label{match1}
    J_0 = + \sqrt{k L_0}\,.
\end{equation}
Periodicity in $\tau$ and triviality around the $\tau$-cycle are equivalent to
the requirement that the holonomy $h$ at $r = 0$ and $r \to \infty$ should lie in the center of the gauge group, i.e.
\begin{equation}
\mathcal{P} \exp  {\int\limits_0^\beta A_{\tau} \dd \tau} \vert_{r = 0/\infty} = \pm \mathbb{1} \qquad \Longleftrightarrow \qquad e^{i \beta V_{0/\infty} \Lt_0} = \pm \mathbb{1}\,
\end{equation}
leading to $\frac{V_0 \beta}{2\pi} = n \in \Z$ and $\frac{V_\infty \beta}{2\pi} =  m \in \Z$. In the following, we impose
\begin{equation}\label{tauperiod}
    n \equiv m = 1 \qquad \Rightarrow \qquad V_0 = V_\infty = \frac {2 \pi} \beta\,.
\end{equation}
Equation \eqref{tauperiod} implies the relation
\begin{equation}
\label{match2}
    \zeta = \mu \sqrt{\frac{4}{k} L_0}\,,
\end{equation}
which using \eqref{BHphimatch} becomes $\zeta = \mu \frac{2}{k} J_0$.

What we have shown above is that the matching conditions relate only the zero modes, $K_0=K_\infty$ and $V_0=V_\infty$. This means that excitations above these zero modes --- e.g.~$\hat u(1)$ excitations at the horizon --- can be switched on arbitrarily and independently. Thus, we see that all soft-hairy horizon descendants at $r = 0$ are compatible with one configuration at $r \to \infty$ described by $\mathcal{L} = \frac{2}{k} L_0$.

\subsection{Gauge connection, metric and asymptotic symmetries}\label{se:3.3}
\label{gaugeconn}

For convenience of the reader we summarize now our results for the first example and translate them also into the metric formulation. The Chern--Simons connection compatible with Brown--Henneaux boundary conditions at $r \to\infty$ and near horizon boundary conditions at $r=0$ is given by
\eq{
A^\pm = b_\pm^{-1}\big(\dd + a^\pm\big)\,b_\pm\,,\qquad\qquad b_\pm = e^{\pm r \Lt_0}\,,\qquad\qquad  a^\pm = g_\pm^{-1} \extd g_\pm\,,
}{eq:angelinajolie}
Inserting the group element
\begin{align}
    g_\pm = e^{\eta^\pm L_{\pm 1}} e^{\pm \psi^\pm L_0}
    e^{\lambda^\pm L_{\mp 1}} e^{\pm r L_0}
\end{align}
yields
\eq{
a^\pm =  e^{\psi^\pm} \dd\eta^\pm \Lt_{\pm 1} \pm \big(\dd\psi^\pm + 2 \lambda^\pm e^{\psi^\pm} \dd\eta^\pm\big)\Lt_0 + \big(\dd\lambda^\pm + \lambda^\pm \dd\psi^\pm + \lambda^{\pm\,2} e^{\psi^\pm} \dd\eta^\pm\big) \Lt_{\mp 1}
}{eq:lalapetz}
and
\begin{align}
\label{eq:bhnh0}
  A^\pm =& e^{\psi^\pm} \dd\eta^\pm e^{r} \Lt_{\pm 1} \pm \big( \extd r +\dd\psi^\pm   + 2 \lambda^\pm e^{\psi^\pm} \dd\eta^\pm\big)
	\Lt_0 \nonumber \\
	&+
	\big(\dd\lambda^\pm + \lambda^\pm \dd\psi^\pm + \lambda^{\pm\,2} e^{\psi^\pm} \dd\eta^\pm\big) e^{- r} \Lt_{\mp 1}\,.
\end{align}
The functions
\begin{align}
\label{eq:bhnh1}
&\eta^\pm = \frac{\eta^\pm_0 + \eta^\pm_\infty\, f(r)}{1 + f(r)}\,,&  &\psi^\pm = \frac{\psi^\pm_0 + \psi^\pm_\infty\, f(r)}{1 + f(r)}\,,&  &\lambda^\pm = \frac{\lambda^\pm_0 + \lambda^\pm_\infty\, f(r)}{1 + f(r)}
\end{align}
depend on both the asymptotic functions $\eta^\pm_\infty$, $\psi^\pm_\infty,\ \lambda^\pm_\infty$, and on the near horizon functions $\eta^\pm_0$, $\psi^\pm_0,\ \lambda^\pm_0$. We choose the interpolating function
\begin{equation}\label{f-sinh}
    f(r) = \sinh^4 r
\end{equation}
to guarantee appropriate fall-off behavior near $r=0$ and $r\to\infty$.

Close to the horizon, $g$ then leads to a connection of the form
\begin{equation}
\label{acorr}
	a = a_0 + {\cal O}(r^4) \dd t + {\cal O}(r^4) \dd\varphi + {\cal O}(r^3) \dd r
\end{equation}
and near infinity we get
\begin{equation}
	a = a_\infty + {\cal O}(e^{-4r}) \dd t + {\cal O}(e^{-4r}) \dd\varphi + {\cal O}(e^{-4r}) \dd r\,.
\end{equation}

For completeness we present also the minus-sector of Brown--Henneaux boundary conditions
\eq{
 A^-_\infty = b\, \big( \extd + a_\infty^- \big)\, b^{-1}\,,
    \qquad\qquad b = e^{r \Lt_0}\,, \qquad\qquad
a_\infty^- =    \big(\Lt_{-1} - \frac{1}{2} \mathcal{L}_- \Lt_{1}\big) \dd \bar{z}\,,
}{connminus}
where we defined $\bar{z} = \varphi - i \mu_- \tau$ and $z = \varphi + i \mu_- \tau$. The on-shell condition 
\begin{equation}
    \partial_{z} {\mathcal{L}}_- = 0
\end{equation}
implies that $\mathcal{L}_- = \mathrm{const}$. In the following we assume that $\mathcal{L}_-$ is positive without loss of generality.  The minus-sector of the near horizon boundary conditions reads
\eq{
 A^-_\infty = b\, \big( \extd + a \big)\, b^{-1}\,,
    \qquad b = e^{r \Lt_0}\,, \qquad
a_0 = - \frac{1}{2} \, \big(\Lt_{-1} - \Lt_{1}\big)\,\big(\mathcal{J}_-(\varphi)\,\extd\varphi - i\zeta_-\,\extd\tau)\,.
}{connminus2}
The minus-sector of the functions \eqref{identlambdainfty} is
\begin{subequations}\label{eq:bhnh2}
\begin{align}
 \eta_\infty^- &= \mp\frac{e^{-\psi_\infty^- }}{\sqrt{2 \mathcal{L}_-}}\,, &  \psi_\infty^- &= \pm \sqrt{2 \mathcal{L}_-} \bar{z}\,, & \lambda_\infty^-  &= \mp \frac{\sqrt{\mathcal{L_-}}}{\sqrt{2}}\,,  \\
\eta_0^- &=\pm \frac{1}{2} e^{- \psi_0^-}\,, &  \psi_0^- &= \pm \int_0^\varphi \mathcal{J}_-(\varphi^\prime) \dd \varphi^\prime \mp i \zeta_- \tau\,, &  \lambda_0^- &= \pm 1\,.
\end{align}	
\end{subequations}
 These functions are subject to the quasi-periodicity conditions \eqref{eq:bhnh02} and the matching conditions\footnote{The modes in the minus-sector are defined in complete analogy to the plus-sector \eqref{decomposition}.}
\begin{equation}\label{matching-zero-modes}
\frac{2}{k}J_0^- = \pm \sqrt{2 \mathcal{L}_-} = \pm \sqrt{\frac{4}{k} L_0^-}\,,\qquad\qquad 
\zeta_- = \pm \mu_- \sqrt{2 \mathcal{L}_-} = \pm \mu_- \sqrt{\frac{4}{k} L_0^-} = \frac{2 \pi}{\beta}\,.
\end{equation}

The metric is obtained from the Chern--Simons gauge field via \cite{Banados:1998gg, Carlip:1998uc}
\eq{
g_{\mu\nu} = \frac{1}{2} \langle (A_\mu^+ - A_\mu^-),(A_\nu^+ - A_\nu^-)\rangle\,.
}{eq:bhnh4}
Inserting \eqref{eq:angelinajolie}-\eqref{eq:bhnh0} yields the metric 
\begin{multline}
\label{eq:bhnh44}
\extd s^2 = \big[ e^{\psi^+} e^r\extd \eta^+  - e^{-r} ( \extd \lambda^- + \lambda^- \extd \psi^- + {\lambda^-}^2 e^{\psi^-} \extd \eta^- ) \big] 
\big[ 
+\leftrightarrow - \big]\\
+ \big( \extd r + \tfrac12\,\big(\extd \psi^+ + \extd \psi^-\big) +  \lambda^+ e^{\psi^+} \extd \eta^+ 
 + \lambda^- e^{\psi^-} \extd \eta^- \big)^2\,.
\end{multline}
The explicit form of  the  near horizon and asymptotic expansions of the metric for the interpolating function $f(r)=\sinh^4 (r)$ is given in appendix \ref{app:B}. 

Here we only show the leading terms in Lorentzian signature ($\tau\rightarrow - i t$). In the asymptotic limit $r\to\infty$ by construction a standard Fefferman--Graham expansion is recovered ($x^\pm =\mu_\pm t \pm \varphi$)
\eq{
\extd s^2\big|_{r\to\infty} = \dd r^2 - e^{2r}\,\dd x^+\dd x^- + \tfrac12\,{\mathcal{L}_+}\,\dd x^{+\,2} + \tfrac12\,{\mathcal{L}_-}\,\dd x^{-\,2} + {\cal O}(e^{-2r})\,.
}{eq:bhnh5}
The near horizon limit $r\to 0$ is compatible with $\mathcal{J}_\pm = \gamma \pm \omega$\,, $\zeta_\pm = -a$ \cite{Afshar:2016wfy, Afshar:2016kjj}
\eq{
\extd s^2\big|_{r\to 0} = \dd r^2 - a^2r^2\,\dd t^2 + \gamma^2\,\dd \varphi^2 \big(1+{\cal O}(r^2)\big) + 2a\omega r^2\,\dd t\dd\varphi + {\cal O}(r^3)\,.
}{eq:bhnh6}
For our matching to work the zero-modes of the state-dependent functions $\mathcal{J}_\pm, \mathcal{L}_\pm$ and the chemical potentials $\zeta_\pm$ and $\mu_\pm$ must be related as in \eqref{match1}, \eqref{match2} and \eqref{matching-zero-modes}. The asymptotic symmetry algebra at $r=0$ consists of two $u(1)$-current algebras \cite{Afshar:2016wfy},
\begin{equation}
\label{algebrasymm ADS}
\big[ J_{n}^{\pm},\,J_{m}^{\pm}\big] = \frac{k}{2}\, n \, \delta_{n+m,\,0}. 
\end{equation}

\subsection{Matching with a Heaviside function}\label{se:3.4}

The discussions in previous sections make it clear that the interpolation function $f(r)$ can be chosen almost arbitrarily, subject only to some fall-off conditions. The specific form \eqref{eq:bhnh1} with some smooth function $f(r)$ is not the only way the interpolation can happen. In particular, we may drop the assumption of smoothness and `interpolate' through a Heaviside function
\begin{align}
\eta^\pm = \eta_0^\pm + \theta(r-r_0) \Delta\eta^\pm, \,\,\;
\psi^\pm = \psi_0^\pm + \theta(r-r_0) \Delta\psi^\pm, \,\,\;
\lambda^\pm = \lambda_0^\pm + \theta(r-r_0) \Delta\lambda^\pm
\end{align}
with some finite critical radius $r_0 > 0$ and the definitions
\begin{align}
    \Delta \eta^\pm = \eta_\infty^\pm - \eta_0^\pm\,,\qquad \qquad
    \Delta \psi^\pm = \psi_\infty^\pm - \psi_0^\pm\,,\qquad \qquad
    \Delta \lambda^\pm = \lambda_\infty^\pm - \lambda_0^\pm\,,
\end{align}
where $\eta_{0/\infty}^\pm, \psi_{0/\infty}^\pm$ and $\lambda_{0/\infty}^\pm$  are given by \eqref{identlambdainfty} and \eqref{eq:bhnh2} and are subject to the matching conditions \eqref{match1}, \eqref{match2} and \eqref{matching-zero-modes}. An advantage of such a choice is that both the asymptotic and near horizon boundary conditions and associated symmetries apply all the way to the critical radius $r_0$, at which point one set of symmetries changes discontinuously to the other.

To deal with $\theta$-function valued functions and their derivatives we use elementary aspects of Colombeau theory, see appendix \ref{app:A} for a derivation of the next formula.
\begin{equation}
\label{colomb}
    f'\big(\theta(r-r_0)\big) \theta'(r-r_0) \sim\big( f(1) - f(0)\big)\,\delta(r-r_0)\,.
\end{equation}
Here $\sim$ stands for equality in the sense of Colombeau. It is then straightforward to write down the explicit form of the Chern--Simons gauge fields with generic value of soft-hairy excitation at the horizon ($r=0$) and a constant Brown--Henneaux excitation at the boundary ($r=\infty$). 

By construction $A_\tau^\pm = A_{\tau,0}^\pm + \theta(r-r_0) ( A_{\tau,\infty}^\pm -A_{\tau,0}^\pm)$ and $A_\varphi^\pm = A_{\varphi,0}^\pm + \theta(r-r_0) ( A_{\varphi,\infty}^\pm -A_{\varphi,0}^\pm)$, while
$A_r^\pm =  \pm \Lt_0 + \delta(r-r_0) {\cal A}^{\pm}_{\mathrm{smooth}}$, for some known functions ${\cal A}^\pm_{\mathrm{smooth}}$. Since the explicit form of the $A_r^\pm$ components is very cumbersome and not illuminating we do not present it here. Restricting the near horizon excitation functions to zero-modes only\footnote{%
We focus for simplicity on the particular matching $\mathcal{J}_\pm = \frac{2}{k} J_0^\pm = \sqrt{2 \mathcal{L}_\pm}$, $\zeta_\pm = \mu_\pm \sqrt{2 \mathcal{L}_\pm}$ explicitly realized by $\eta_0^\pm = \frac{1}{2} \exp(- \psi_0^\pm),\, \psi_0^\pm = \mathcal{J}_\pm \varphi \pm i \zeta_\pm \tau,\, \lambda_0^\pm = 1$\,, $\eta_\infty^+ = - \exp(- \psi_\infty^+)/\sqrt{2\mathcal{L}_+}, \, \psi_\infty^+ = \sqrt{2 \mathcal{L}_+} z, \,\lambda_\infty^+ = - \sqrt{\mathcal{L_+}/2}$ and $\eta_\infty^- = - \exp(- \psi_\infty^-)/\sqrt{2\mathcal{L}_+}, \, \psi_\infty^- = \sqrt{2 \mathcal{L}_-} \bar{z}, \lambda_\infty^- = - \sqrt{\mathcal{L_-}/2}$, where $z = i \mu_+ \tau + \varphi$ and $\bar{z} = - i \mu_- \tau + \varphi$.}
simplifies the radial component of the connection, 
\begin{align}
 \label{constdist}
A_r^+ 
=& - e^r\Big(\frac{1}{\mathcal{J}_+}+\frac{1}{2} \Big)\,\delta(r-r_0) \Lt_1+  \Big(\frac{1}{4}\mathcal{J}_+ -\frac{1}{\mathcal{J}_+} \Big)\, \delta(r-r_0) \Lt_0 + \Lt_0\nonumber \\
&-e^{-r}\Big(\frac{1}{24} \mathcal{J}_+^2 +\frac{1}{2} \mathcal{J}_+ +\frac{1}{3 \mathcal{J}_+}+1 \Big)\,\delta(r-r_0) \Lt_{-1} \,.
\end{align}
The connection component $A_r^-$ is obtained from $A_r^+$ by exchanging $\Lt_1 \leftrightarrow \Lt_{-1}, \Lt_0 \leftrightarrow - \Lt_0 , \mathcal{J}_+  \leftrightarrow \mathcal{J}_-$. 

We end this subsection with the remark that despite of having $\theta$- and $\delta$-functions in the gauge fields $A_\mu^{\pm}$, the associated field strengths (by construction) are zero everywhere. Thus, in the metric formulation the spacetime curvature is locally AdS$_3$ everywhere, even though the metric component $g_{rr}$ contains terms quadratic in the $\delta$-function.

\section{Comp{\`ere}--Song--Strominger/Heisenberg matching}\label{se:4}

As our second example we consider the matching of near horizon boundary conditions at $r = 0$ to Comp{\`e}re--Song--Strominger (CSS) boundary conditions \cite{Compere:2013bya} at $r = \infty$. In this section we present both the $A^+$ and the $A^-$ sector, since the sectors are qualitatively different. The connection subject to CSS boundary conditions at $r \to \infty$ in Euclidean signature reads
\begin{subequations}
\label{cpssbc}
\begin{align}
    A^\pm &= b^{\mp 1}\,\big(\extd+a^\pm\big)\, b^{\pm 1}\,,\qquad \qquad b = e^{r\Lt_0}\,, \\
a^+ &= \big(- \partial_z P \extd z - \extd\bar{z}\big) \Big(\Lt_1 - \frac{\Delta}{k}\, \Lt_{-1}\Big)\,, \\
a^- &= \Big( \frac{L}{2} \Lt_1 - \Lt_{-1}\Big) \extd z\,,
\end{align}
\end{subequations}
where $z = i \tau + \varphi$ and $\bar{z} = - i \tau + \varphi$ and $\partial_{\bar{z}} P = 0$. Here, $\Delta$ is a constant while $P(z, \bar{z})$ and $L(z,\bar{z})$ are subject to the on-shell conditions
\begin{equation}
   \partial_{\bar{z}} \partial_z P = 0\,,\qquad  \qquad \partial_{\bar{z}} L = 0\,.
\end{equation}
As before, this implies that $L$ and $\partial_z P$ are constants. This yields
\begin{equation}
    P(z, \bar{z}) = P_0 z+P_1\,, \qquad\qquad L(z)=L\, .
\end{equation}

In the following for simplicity and without loss of generality we assume that $\Delta$ and $L$ are positive. The functions $\eta^\pm, \psi^\pm$ and $\lambda^\pm$ for the group element $g$ read  
\begin{align}
    \lambda_\infty^+ &= \pm \sqrt{\frac{\Delta}{k}}\,,  &  \psi_\infty^+ &= \pm 2 \sqrt{\frac{\Delta}{k}} (\bar{z} + P_0z+P_1)\,, & \eta_\infty^+ &= \pm \sqrt{\frac{k}{\Delta}} \frac{e^{- \psi_\infty^+}}{2}\,, \\
    \lambda_\infty^- &= \pm \sqrt{\frac{L}{2}}\,,   &
    \psi_\infty^- &= \pm \sqrt{2  L} z\,,  &
    \eta_\infty^- &=\pm \frac{e^{- \psi_\infty^-}}{\sqrt{2 L}} \,.
\end{align}
The solution satisfies the periodicity conditions \eqref{eq:bhnh02} and \eqref{eq:bhnh3} with
\begin{equation}
K_\infty^+ =  \pm 2 \sqrt{\frac{\Delta}{k}} (P_0 + 1)\,, \qquad V_\infty^+ = \pm 2 \sqrt{\frac{\Delta}{k}} (P_0 - 1)\,,\qquad K_\infty^- = V_\infty^- = \pm \sqrt{2 L}\,.   \end{equation}
The matching conditions analogous to section \ref{se:3.2} read $V_\infty^\pm = V_0^\pm = \frac{2 \pi}{\beta}$ and $K_\infty^\pm = K_0^\pm$. Using the results from sections \ref{conn2boundBH} and \ref{gaugeconn} yields  
\begin{subequations}
\begin{align}
\zeta^+ &=  \pm 2 \sqrt{\frac{\Delta}{k}} (P_0 - 1) = \frac{2 \pi}{\beta}\,, & \zeta^- &=  \pm \sqrt{2 L}= \frac{2 \pi}{\beta}\,,\\ 
\frac{2}{k} J_0^+ &= \pm 2 \sqrt{\frac{\Delta}{k}} (P_0 + 1)\,, & \frac{2}{k} J_0^- &= \pm \sqrt{2 L}\,.
\end{align}    
\end{subequations}

In summary, we can smoothly flow from the CSS boundary conditions at infinity with fixed (zero-mode) excitations to the Heisenberg boundary conditions at the horizon with  generic excitations. Thus, also black holes in AdS$_3$ with CSS boundary conditions can be equipped with arbitrary soft hair excitations.

\section{Concluding remarks}\label{se:5}

We showed how to interpolate between two different sets of boundary conditions in AdS$_3$ Einstein gravity using the Chern--Simons formulation. The key result of section \ref{se:2} is that {with our procedure} any two sets of boundary conditions can be interpolated as long as the matching conditions hold. In particular, the holonomies around both cycles should yield the same when evaluated near either of the boundaries. 

In our two examples, Brown--Henneaux/Heisenberg matching in section \ref{se:3} and Comp\`ere--Song--Strominger/Heisenberg matching in section \ref{se:4}, we interpolated between the asymptotic region and the near horizon region. As we saw in both examples, a physical consequence of the matching conditions is that the mass and angular momentum at infinity must match with corresponding charges near the horizon. However, the tower of Fourier modes of excitations at the horizon is totally unconstrained by this matching, so that any amount of soft hair excitations is allowed for any given values of mass and angular momentum, as measured by an asymptotic observer. 

We conclude now with a list of further possible applications and generalizations.
\begin{itemize}
    \item {\bf Lorentzian signature.} Our Euclidean results generalize to Lorentzian signature. Reconsider our prime example of section \ref{se:3}, but with Lorentzian signature. In that case classical solutions can have generic Brown--Henneaux excitations asymptotically, specified by two arbitrary periodic functions $\mathcal{L}_\pm(x^\pm)$. Nonetheless, the matching conditions relate only the zero modes of the Brown--Henneaux and Heisenberg charges, while all other boundary gravitons can be excited independently at either of the boundaries. This dovetails with the detailed analysis of  Ba{\~n}ados geometries \cite{Banados:1998gg} and that the entropy and angular momentum are orbit invariant quantities \cite{Sheikh-Jabbari:2016unm}.
    \item {\bf Auto-interpolation.} It is of course possible to impose the same set of boundary conditions at either of the boundaries. For instance, one can impose Brown--Henneaux boundary conditions at the conformal boundary of AdS$_3$ and at an arbitrary constant radius-slice, e.g.~a stretched horizon. We do not present the details of the analysis here but mention the final result for the Lorentzian case. Starting with a Ba{\~n}ados geometry \cite{Banados:1998gg} near one boundary one may flow to another Ba{\~n}ados geometry at the other boundary and the matching conditions tell us that the zero modes must not change. In more algebraic wording, recall that the Ba{\~n}ados geometries are in one-to-one relation with Virasoro coadjoint orbits \cite{Sheikh-Jabbari:2016unm}. Our matching procedure then means that one can flow within a given coadjoint orbit from one representant to another, but cannot move across the orbits. 
    \item {\bf Flow of boundary action.} While our focus was on the interpolation of solutions with different boundary conditions, it should be possible to interpolate between corresponding boundary actions. For instance, the Brown--Henneaux boundary conditions lead to a Liouville-like boundary theory \cite{Coussaert:1995zp, Barnich:2017jgw, Cotler:2018zff}, which near the horizon should flow to a Cangemi--Jackiw type of scalar field theory \cite{Grumiller:2019tyl}. It would be nice to verify this and also to generalize it to flows between other sets of boundary actions.
    \item {\bf Fluff proposal.} An application of soft hair excitations \cite{Hawking:2016msc} is the fluff proposal \cite{Afshar:2016uax,Afshar:2017okz, Sheikh-Jabbari:2016npa}, which relies on several working assumptions. One of them is that there are non-trivial physical boundary excitations near the horizon that are invisible to an asymptotic observer, which is precisely what we have shown in the present work. We hope to address the remaining working assumptions in the future, in particular a mechanism that provides a cut-off on the soft hair spectrum.
    \item {\bf Higher spin and asymptotic flat theories.} Our results generalize straightforwardly to other gravity or gravity-like theories, like higher spin gravity \cite{Henneaux:2010xg, Campoleoni:2010zq} or asymptotically flat gravity \cite{Barnich:2006av} or both \cite{Afshar:2013vka, Gonzalez:2013oaa}, as long as there is a Chern--Simons formulation. In particular, higher spin black holes \cite{Gutperle:2011kf} with Brown--Henneaux-like boundary conditions (or their flat space cosmology cousins \cite{Gary:2014ppa, Matulich:2014hea}) can again be interpolated to near horizon Heisenberg boundary conditions \cite{Grumiller:2016kcp, Ammon:2017vwt}.
    \item {\bf Higher derivative theories.} It could be of interest to generalize our discussion to higher derivative theories of gravity, which were studied vigorously in the past decade, see e.g.~\cite{Hinterbichler:2011tt}. In three dimensions a large class of these theories allows a formulation as Chern--Simons-like theories \cite{Merbis:2014vja, Bergshoeff:2014bia}. It seems likely that most of our conclusions carry over to such theories. 
    \item {\bf Higher dimensional theories.} While our technical implementation of the interpolation relied on the Chern--Simons formulation, we could have done everything in the metric formulation. On general grounds, we expect that it must be possible to interpolate between different sets of boundary conditions also in higher dimensions, though it is unclear to us what the precise matching conditions will be. For instance, interpolating between asymptotically AdS$_D$ boundary conditions (see e.g.~\cite{Skenderis:2002wp}) and near horizon boundary conditions \cite{Grumiller:2019fmp} will be rewarding to better understand non-extremal black holes and their soft hairy counterparts.
    \item {\bf Non-gravitational applications.} Since  our technical implementation of the interpolation relied on the Chern--Simons formulation, our results apply also to non-gravitational applications of Chern--Simons in presence of two boundaries. A potential application is a Quantum-Hall system with two disconnected boundary components and different spectra of edge excitations at either of the boundaries.
    \item {\bf More boundaries.} Finally, it seems interesting to generalize the discussion to more than two disconnected boundary components. For instance, the recent holographic model for black hole evaporation \cite{Akers:2019nfi} involves configurations with three and more boundary components, and it could be worthwhile to impose boundary conditions separately on each of these components.
\end{itemize}
We intend to come back to some of these generalizations in the future.

\section*{Acknowledgements}

We are grateful to Hamid Afshar,  Wout Merbis, Mateusz Piorkowski, and Friedrich Sch\"oller for useful discussions. 

DG was supported by the Austrian Science Fund (FWF), projects P~28751, P~30822 and P~32581. MMShJ would like to thank the hospitality of ICTP HECAP where this work finished. He also acknowledges the support by INSF grant No 950124 and Saramadan grant No. ISEF/M/98204. RW was supported by a DOC Fellowship of the Austrian Academy of Sciences and by the Austrian Science Fund FWF under the Doctoral Program W1252-N27 Particles and Interactions. DG and MMShJ acknowledge the Iran-Austria IMPULSE project grant, supported and run by Khawrizmi University. During part of this work, CT was supported by the Max Planck Institute for Gravitational Physics (Albert Einstein Institute) in Potsdam.

\appendix

\section{Aspects of Colombeau theory}
\label{app:A}

Colombeau theory gives a mathematical framework of multiplying distributions \cite{colombeau1990}. In a general relativistic context the need for Colombeau theory can arise when considering distributional metrics \cite{Clarke:1996pp, Steinbauer:1996fv, Balasin:1996mq, Steinbauer:2006qi}, e.g.~for solutions of Einstein gravity describing gravitational shock-waves (though often Colombeau-theory can be avoided by choosing suitable coordinates).

In our work we are interested in making sense of first derivatives of smooth functions of the Heaviside function. The essence of Colombeau theory applied to this case is to replace the Heaviside function by some smooth family of functions, $\theta_\epsilon$, that in the limit of vanishing $\epsilon$ yields the Heaviside function $\theta$, and to manipulate the expressions in such a way that the final result does not depend on the precise family.

The result \eqref{colomb} then follows from a chain of equalities
\begin{multline}
    \lim_{\epsilon \rightarrow 0} \int\limits_{-\infty}^\infty\! f'(\theta_\epsilon(x)) \theta_\epsilon'(x) \psi(x) \extd x
    = - \lim_{\epsilon \rightarrow 0} \int\limits_{-\infty}^\infty\! f(\theta_\epsilon(x)) \psi'(x) \extd x 
     \\ = - \lim_{\epsilon \rightarrow 0} \int\limits_{-\infty}^\infty (f(\theta_\epsilon(x)) - f(0)) \psi'(x) \extd x 
     = - \int\limits_0^\infty \left(f(1)- f(0)\right) \psi'(x) \extd x \\ = 
    \left(f(1)- f(0)\right) \psi(0) =  \left(f(1)- f(0)\right) \int\limits_{-\infty}^\infty \delta(x) \psi(x) \extd x\,,
    \label{eq:colo1}
\end{multline}
where $\psi(x)$ is a test-function with the usual properties. The result \eqref{eq:colo1} is independent of the approximation used for the Heaviside function and allows to associate $f'(\theta(x))\,\theta'(x)$ to the distribution $\delta(x)$ times a prefactor that we just derived
\begin{equation}
    f'(\theta(x)) \theta'(x) \sim \left(f(1)- f(0)\right)\, \delta(x)\,.
\end{equation}
The choice $f'(\theta(x)) = \theta^n(x)$ recovers the classic example by Colombeau \cite{colombeau1990}
\begin{equation}
    \theta^n(x) \theta'(x) \sim \frac{1}{n+1}\, \delta(x)\,.
\end{equation}

\section{Explicit form of the interpolating metric}\label{app:B}

In this appendix we provide the near horizon and the asymptotic expansion 
of the metric that interpolates between Brown Henneaux boundary conditions at infinity and Heisenberg boundary conditions at $r= 0$. We display our results in Lorentzian signature and choose our connection such that $\frac{2}{k} J_0^\pm =  \sqrt{2 \mathcal{L}_\pm}$ and  $\zeta_\pm =  \mu_\pm \sqrt{2 \mathcal{L}_\pm} = -a$, which is explicitly realized by
 \begin{subequations}
 \label{eq:bhnh323}
\begin{align}
 \eta_\infty^+ &= - \frac{e^{- \psi_\infty^+}}{\sqrt{2 \mathcal{L}_+}}\,, & \psi_\infty^+ &=  \sqrt{2 \mathcal{L}_+} (\varphi + \mu_+ t)\,, & \lambda_\infty^+  &=  - \frac{\sqrt{\mathcal{L}_+}}{\sqrt{2}}\,,  \\
\eta_0^+ &=  \frac{1}{2} e^{- \psi_0^+}\,, & \psi_0^+ &=  - a t + \int_0^\varphi (\gamma(\varphi^\prime) + \omega(\varphi^\prime))\dd \varphi^\prime\,, & \lambda_0^+ &= 1\,, \\
 \eta_\infty^- &= -\frac{e^{- \psi_\infty^-}}{\sqrt{2 \mathcal{L}_-}}\,, &  \psi_\infty^- &=  \sqrt{2 \mathcal{L}_-} (\varphi - \mu_- t)\,, & \lambda_\infty^-  &= - \frac{\sqrt{\mathcal{L_-}}}{\sqrt{2}}\,, \\
\eta_0^- &= \frac{1}{2} e^{- \psi_0^-}\,, & \psi_0^- &=   a t + \int_0^\varphi (\gamma(\varphi^\prime) - \omega(\varphi^\prime))\dd \varphi^\prime\,, & \lambda_0^- &=  1\,.
\end{align}
\end{subequations}

The interpolating function $f(r) = \sinh^4(r)$ together with \eqref{eq:bhnh3} after Taylor expanding around $r = 0$ yields
\begin{subequations}
    \label{eq:bhnh441}
    \begin{align}
        g_{tt} &= -a^2 r^2-\frac{a^2 r^4}{3} + O(r^5) \\
      g_{\varphi \varphi} &= \gamma^2(\varphi)+r^2 \left(\gamma^2(\varphi)-\omega^2(\varphi)\right)+ r^4 \biggl(
           \gamma (\varphi )
      \left(\sqrt{2 \mathcal{L}_-}+\sqrt{2 \mathcal{L}_+}\right)
       -5 \frac{\gamma^2(\varphi)}{3}
      - \frac{\omega^2(\varphi)}{3}\biggr) \nonumber \\
      & \quad + O(r^5)\\
      g_{rr} &= 1+r^3 \left(8
      \left(-\frac{e^{\tilde{\mathcal{J}}_-(\varphi
      )}}{\sqrt{2\mathcal{L}_-}}-\frac{e^{\tilde{\mathcal{J}}_+(\varphi
      )}}{\sqrt{2\mathcal{L}_+}}\right)-4
      \tilde{\mathcal{J}}_-(\varphi )-4
      \tilde{\mathcal{J}}_+(\varphi
      )-8\right)+O\left(r^5\right) \\
      g_{\varphi r} &=-r^3 \gamma (\varphi ) \left(2
      \tilde{\mathcal{J}}_-(\varphi )+2
      \tilde{\mathcal{J}}_+(\varphi )+\sqrt{2 \mathcal{L}_-}+\sqrt{2 \mathcal{L}_+}+4\right)+O\left(r^4\right) \\
       g_{tr}&= \frac{3}{4} a r^4 \left( \left(-\frac{4
       e^{\tilde{\mathcal{J}}_-(\varphi
       )}}{\sqrt{2 \mathcal{L}_-}}+\frac{4
       e^{\tilde{\mathcal{J}}_+(\varphi
       )}}{\sqrt{2\mathcal{L}_+}}-\sqrt{2\mathcal{L}_-}+\sqrt{2\mathcal{L}_+}\right)-2
       \tilde{\mathcal{J}}_-(\varphi )+2
       \tilde{\mathcal{J}}_+(\varphi )\right)+O\left(r^5\right)\\
       g_{t\varphi} &= a r^2 \omega(\varphi) +\frac{1}{3} a r^4 \omega(\varphi) +O\left(r^5\right)\,,
    \end{align}
    \end{subequations}
    where $\tilde{\mathcal{J}}_\pm(\varphi) = - \sqrt{2 \mathcal{L}_\pm} + \int_0^\varphi \mathcal{J}_\pm(\varphi^\prime) 
    \dd \varphi^\prime = \int_0^\varphi \frac{2}{k} \sum_{n \neq 0} J_n^\pm e^{-i n \varphi^\prime} \dd \varphi^\prime$.
    We see that $r=0$ is a Killing horizon of the vector $\xi = \partial_t$. Expanding around $r \to \infty$ yields
    \begin{subequations}
        \label{eq:bhnh42}
        \begin{align}
            g_{tt} &= 
        -\mu_- \mu_+ e^{2r}+\frac{1}{2} \left(\mu_-^2 \mathcal{L}_- +\mu_+^2
           \mathcal{L}_+\right) + \frac{1}{4} \mu_- \mu_+ e^{-2 r} \biggl(-64
           \tilde{\mathcal{J}}_-(\varphi )-64
           \tilde{\mathcal{J}}_+(\varphi )-\mathcal{L}_- \mathcal{L}_+ \nonumber \\
           &\quad +32 \sqrt{2 \mathcal{L}_-}
           e^{-\tilde{\mathcal{J}}_-(\varphi )}+32 \sqrt{2 \mathcal{L}_+}
           e^{-\tilde{\mathcal{J}}_+(\varphi
           )} +128\biggr)+O\left(e^{- 4r}\right)\\
           g_{\varphi \varphi} &= e^{2 r} +\frac{1}{2} \left(\mathcal{L}_-+\mathcal{L}_+\right)+ \frac{1}{4} e^{-2 r}
            \biggl(32
           e^{-\tilde{\mathcal{J}}_-(\varphi )} (\omega (\varphi
           )-\gamma (\varphi ))-32 e^{-\tilde{\mathcal{J}}_+(\varphi )} (\gamma
           (\varphi )+\omega (\varphi ))\nonumber \\
           &\quad +64 \tilde{\mathcal{J}}_-(\varphi
           )+64
           \tilde{\mathcal{J}}_+(\varphi )+\mathcal{L}_-
           \mathcal{L}_+-128\biggr)
           +O\left(e^{-4r}\right) \\
          g_{rr} &=  1+O\left(e^{- 4r}\right)\\
          g_{\varphi r} &= 16 e^{-2 r} \biggl(\sinh (\tilde{\mathcal{J}}_-(\varphi
          ))-\cosh (\tilde{\mathcal{J}}_-(\varphi ))+\sinh
          (\tilde{\mathcal{J}}_+(\varphi ))-\cosh
          (\tilde{\mathcal{J}}_+(\varphi ))\nonumber \\
          &\quad+\sqrt{2}
          \left(-\frac{1}{\sqrt{\mathcal{L}_-}}-\frac{1}{\sqrt{\mathcal{L}_+}}\right)\biggr) +O\left(e^{- 4r}\right)
           \\
           g_{t r} & = 16 e^{-2 r} \left(-\mu_+
           e^{-\tilde{\mathcal{J}}_-(\varphi )}+\mu_-
           e^{-\tilde{\mathcal{J}}_+(\varphi )}+\sqrt{2}
           \left(\frac{\mu_-}{\sqrt{\mathcal{L}_+}}-\frac{\mu_+}{\sqrt{\mathcal{L}_-}}\right)\right) +O\left(e^{- 4r}\right) \\
           g_{t \varphi} &= \frac{\mu_+-\mu_-}{2} e^{2r}+\frac{1}{2} \left(\mu_+ \mathcal{L}_+ -\mu_-
           \mathcal{L}_-\right)
           + e^{-2 r} \biggl(4 \mu_+
   e^{-\tilde{\mathcal{J}}_-(\varphi )}( \omega (\varphi )-\gamma (\varphi )) \nonumber \\
   &\quad-8
  ( \tilde{\mathcal{J}}_+(\varphi ) + \tilde{\mathcal{J}}_-(\varphi )) (\mu_- -\mu_+)+4 \mu_- \sqrt{2 \mathcal{L}_-}
   e^{-\tilde{\mathcal{J}}_-(\varphi )}+4 \mu_-
   e^{-\tilde{\mathcal{J}}_+(\varphi )} (\gamma(\varphi) +\omega (\varphi ))\nonumber \\
   &\quad -4
    \mu_+ \sqrt{2 \mathcal{L}_+}
   e^{-\tilde{\mathcal{J}}_+(\varphi )}-16 (\mu_+ - \mu_-) +\frac{(\mu_+ - \mu_-)
   \mathcal{L}_- \mathcal{L}_+}{8}\biggr)+O\left(e^{- 4r}\right)\,.
        \end{align}
        \end{subequations}
         
Restricting the solution to zero modes $\mathcal{J}_\pm = \frac{2}{k} J_0^\pm = \gamma \pm \omega$ ($\tilde{\mathcal{J}}_\pm(\varphi)=0$) the near horizon expansion \eqref{eq:bhnh441} reduces to
\begin{subequations}
    \label{eq:bhnh41}
\begin{align}
    g_{tt} &= -a^2 r^2-\frac{a^2 r^4}{3} + O(r^5) \\
  g_{\varphi \varphi} &= \gamma ^2+r^2 \left(\gamma ^2-\omega ^2\right)+\frac{1}{3} r^4 \left(\gamma ^2-\omega ^2\right) + O(r^5)\\
  g_{rr} &= 1 - r^3 \left(\frac{8}{\gamma
   +\omega }-\frac{8}{\omega
   -\gamma
   }+8\right)+O\left(r^5\right) \\
  g_{\varphi r} &= -2 (\gamma  (\gamma +2)) r^3+O\left(r^4
   \right) \\ 
   g_{tr}&= \frac{3 a r^4 \left(\gamma ^2 \omega -\omega ^3-4 \omega \right)}{2 (\gamma -\omega ) (\gamma +\omega )}+O\left(r^5\right)\\
   g_{t\varphi} &= a r^2 \omega +\frac{1}{3} a r^4 \omega +O\left(r^5\right)\,.
\end{align}
\end{subequations}
and the asymptotic expansion \eqref{eq:bhnh42} reduces to
\begin{subequations}
\begin{align}
\label{eq:bhnh432}
    g_{tt} &= 
-\mu_- \mu_+ e^{2r}+\frac{1}{2} \left(\mu_-^2 \mathcal{L}_- +\mu_+^2
   \mathcal{L}_+\right) + \frac{1}{4} \mu_- 
   \mu_+  e^{-2 r}
   \biggl(-\mathcal{L}_-
   \mathcal{L}_++32
   \sqrt{2 \mathcal{L}_-} \nonumber \\
   &\quad
   + 32 \left(\sqrt{2 \mathcal{L}_+}+
   4\right)\biggr) +O\left(e^{- 4r}\right)\\
   g_{\varphi \varphi} &= e^{2 r} +\frac{1}{2} \left(\mathcal{L}_-+\mathcal{L}_+\right)+ \frac{1}{4} e^{-2 r} \left(\mathcal{L}_- \mathcal{L}_+-32 \sqrt{2}
   (\sqrt{\mathcal{L}_-}+\sqrt{\mathcal{L}_+})-128 \right) +O\left(e^{-4r}\right) \\
  g_{rr} &=  1+O\left(e^{- 4r}\right)\\
  g_{\varphi r} &= e^{-2 r} \left(-\frac{16 \sqrt{2}}{\sqrt{\mathcal{L}_-}}-\frac{16
   \sqrt{2}}{\sqrt{\mathcal{L}_+}}-32\right) +O\left(e^{- 4r}\right)\\
   g_{t r} & = 16 e^{-2 r} \left(\mu_- -\mu_+
   \left(\frac{\sqrt{2}}{\sqrt{\mathcal{L}_- }}+1\right)+\frac{\sqrt{2}
   \mu_-}{\sqrt{\mathcal{L}_+ }}\right
   )  +O\left(e^{- 4r}\right) \\
   g_{t \varphi} &= \frac{\mu_+-\mu_-}{2} e^{2r}+\frac{1}{2} \left(\mu_+ \mathcal{L}_+ -\mu_-
   \mathcal{L}_-\right) + \frac{1}{8} e^{-2 r} (\mu_- -\mu_+ ) \biggl(-\mathcal{L}_-
   \mathcal{L}_++32 \sqrt{2 \mathcal{L}_-}\nonumber \\
   &\quad+32 \left(\sqrt{2 \mathcal{L}_+}+4\right)\biggr) +O\left(e^{- 4r}\right)\,.\end{align}
\end{subequations}
From this last set of equations it is clear that we have a usual Fefferman--Graham expansion at large radii, namely a fixed metric to order ${\cal O}(e^{2r})$ and state-dependent contributions to ${\cal O}(1)$. However, to subleading order ${\cal O}(e^{-2r})$ we deviate from the standard form of asymptotically AdS$_3$ solutions due to the various square-root terms, which are an echo of the near horizon structure that asymptotically is perceived as pure gauge.

\bibliographystyle{fullsort}
\addcontentsline{toc}{section}{References}
\bibliography{review}

\end{document}